\documentclass[aps,twocolumn,showpacs]{revtex4}

\usepackage{graphicx}

\def\fig#1{Fig.\,\ref{#1}}

\def\eq#1{Eq.~(\ref{#1})}

\def\dr{\langle\Delta r\rangle}

\begin{document}
\title{Ionization and charge migration through strong internal
  fields \\ in clusters exposed to intense X-ray pulses}
\author{Christian Gnodtke, Ulf Saalmann and Jan M. Rost}
\affiliation{Max Planck Institute for the 
  Physics of Complex Systems,
  N\"othnitzer Stra{\ss}e 38, 01187 Dresden, Germany \\
  and  Max Planck Advanced Study Group at the
  Center for Free-Electron Laser Science, Hamburg, Germany}
\date{\today}

\begin{abstract}\noindent 
  A general scenario for electronic charge migration in 
  finite samples illuminated by an intense laser pulse is given.
  Microscopic calculations for neon clusters under strong short
  pulses as produced by X-ray free-electron laser sources
  confirm this scenario and point to the prominent role of field
  ionization by strong internal fields.  
  The latter leads to the fast formation of a core-shell 
  system with an almost static core of screened ions while the
  outer shell explodes. 
  Substituting the shell ions with a different material such as
  helium as a sacrificial layer leads to a substantial
  improvement of the diffraction image for the embedded cluster
  thus reducing the consequences of radiation damage for
  coherent diffractive imaging.  
\end{abstract}


\maketitle

\noindent The advent of sources producing short and intense
pulses of light with frequencies ranging from soft to hard
X-rays \cite{fear+05} opens a new parameter regime for
light-matter interaction.   
Dynamics with a quick removal of many electrons from their bound
states in the atoms of the sample irradiated has scarcely been
explored before.  
It is important to have a good insight into this dynamics in
order to realize one of the goals of X-ray free electron lasers
(XFELs) which deliver such light pulses: 
the single-shot coherent diffractive imaging of finite
samples with atomic resolution \cite{gach07}.  
X-ray photons scatter elastically from a single
non-periodic molecule, like a protein, and  are subsequently
recorded on a CCD as a continuous diffraction pattern \cite{newo+00}.
The estimated total photon flux required for single-shot diffraction
imaging exceeds $10^{12}$ photons per (100\,nm)$^2$
\cite{newo+00,halo+05,miho+01}, and is at the limit of what
XFELs can deliver.   
Above all,  such high photon fluxes lead to radiation damage by
multiple ionization and subsequent Coulomb explosion of the sample 
which degrades the diffraction image if not destroys it.

\begin{figure}[b]
\includegraphics[width=0.32\textwidth]{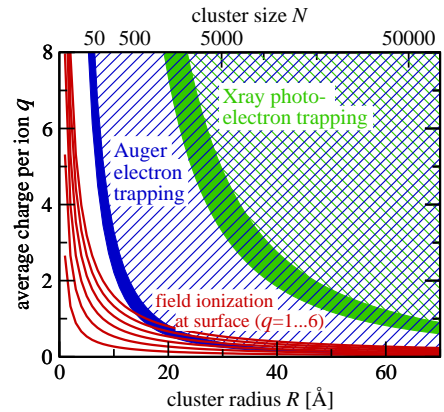}
\caption{(color online.) Electron trapping and strong-field effects of
  homogeneously charged neon  clusters with an average charge
  $q$ per ion and a cluster radius $R$. 
  Trapping of electrons detached by 12\,keV photons
  (green-hatched region) and Auger decays (blue-hatched region). 
  The thick lower border (dark-green and dark-blue bands) take
  into account that electrons are released from any position in
  the cluster.  
  Red lines: Ionization of surface ions at various charge
  states due to the internal radial field of the charged cluster
  estimated by the Bethe rule, see text.}
\label{fig:elcapture}
\end{figure}%

For possible means to overcome this handicap a detailed
understanding of the time-dependent multi-particle dynamics is
required which is very interesting in its own right. 
Apart from simulations in the X-ray  regime
\cite{newo+00,saro02,jufa+04,juos+04,halo+04,halo+07,hach08},  
this problem has been approached from the perspective of intense
long-wavelength laser pulses.  
There, it was found that finite systems, such as atomic clusters
or nano-droplets, display  behavior fundamentally different from
single atoms or bulk systems \cite{krsm02,sasi+06}. 
The large space charge of ions in a finite volume creates strong
electric fields easily exceeding the strong external field of the laser. 
Similarly, very strong internal fields are generated by
single-photon ionization in X-ray pulses since many photo
electrons can leave the sample due to the high photon energy. 
In both cases the strong internal field leads to the release of
additional electrons, mainly from the surface.  
In the X-ray regime this effect has profound consequences
for diffractive imaging, as we will show. 

As long as the laser frequency is sufficiently high to trigger
initial ionization, a unified scenario with four phases emerges
for a sample illuminated by an intense light pulse: 
i)~Electrons are photo-detached from the atoms by the laser and
leave the cluster.  
ii)~The substantial ionic charge developed during
the first phase generates a large electric field particularly at
the cluster  surface, leading to almost instantaneous field
ionization of surface atoms.   
These electrons migrate towards the cluster center.  
iii)~An electron plasma is established, initially formed
by field-ionized electrons.
For sufficiently large samples secondary electrons (from Auger
decay or electron-impact ionization) and even photo-detached 
electrons are trapped by the cluster potential, ``feeding''
the plasma. 
iv)~The electron plasma shields the core of the cluster
so that only the (highly charged) outer shell explodes.

This scenario occurs for many combinations of parameters
(sample size and density, pulse length, frequency and intensity)
provided that the quiver amplitude $x = F/\omega^{2}$, i.e., the
excursion of electrons induced by the oscillating laser field
with amplitude $F$ and frequency $\omega$, is much smaller than
the characteristic size (linear extension $X$) of the sample,
$x\ll X$. 
In the X-ray regime \cite{jufa+04,halo+04} this is fulfilled even for high field
strengths, but it has been shown to occur also for
much smaller frequencies in the VUV \cite{siro04}.
Recently, this electron migration has been elegantly verified
experimentally with an ad hoc designed core-shell system of a
xenon-argon composite cluster \cite{hobo+08}. 

Figure~\ref{fig:elcapture} shows at which
combination of cluster radius and average charge per ion trapping of photo-
and Auger electrons is expected  for neon clusters exposed to 
a 12 keV photon pulse, relevant for atomic-resolution imaging. 
One sees immediately, that large clusters  trap even X-ray
photo-electrons while samples of moderate size trap only the much slower
Auger electrons.  
Field ionization, in contrast, occurs from the smallest sample
sizes on.  
What contributes even more to their prominent role for  the
dynamical screening of ions is the fact that the field ionized
electrons are not only available in a sample of arbitrary size
but they are also the first electrons available for screening
once the ionizing laser pulse sets in. 
Auger electrons, on the other hand, appear only about 2.5\,fs
after a K-shell photo-ionization event in neon, when the core
hole undergoes an Auger decay.

The critical field ${\cal E}_\mathrm{c}$ for
ionization of an atom  can be estimated by the Bethe-rule
\cite{besa57} ${\cal E}_\mathrm{c} = E_\mathrm{ip}{}^2/(4Z)$, 
where $E_\mathrm{ip}$ is the ionization potential and $Z$ is
the charge of the binding core.  Hence, fields produced in neon
clusters as small as $R \approx 10$\,{\AA} are sufficient for multiple
ionization of the surface atoms (red lines in \fig{fig:elcapture})
while larger clusters with radii $R \gtrsim 50$\,{\AA} reach the
critical field strength at the surface with only a small fraction of
photo-ionized atoms.  This suggests field ionization as an extremely
efficient and fast mechanism for an electron migration that leads to
screening of the core of the sample.

\begin{figure}[t]
\includegraphics[width=0.5\textwidth]{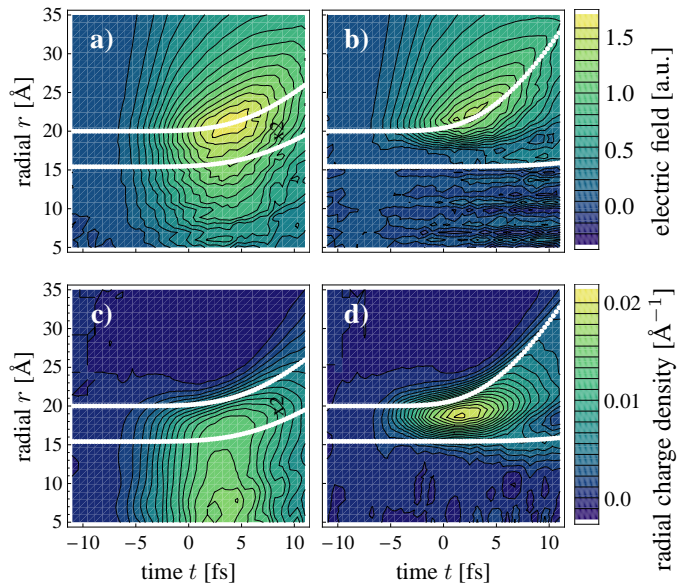}
\caption{(color online.) Radial electric field (upper panels) and radial
  charge density (lower panels) for a $\text{Ne}_{1500}$ cluster
  and $T=10$\,fs pulse 
  without the effect of field-ionization (left) and with field
  ionization (right). 
  White lines show the average radial coordinate of cluster
  surface shell (the 5\% outermost ions) and of cluster-core
  surface shell (the 5\% outermost ions of the inner half of the
  cluster).}
\label{fig:contours}
\end{figure}%
To assess the effect of field ionization quantitatively we have
calculated the dynamics of neon clusters of $N=50$
($R\approx6$\,\AA) up to $N=15000$ ($R\approx42$\,\AA) atoms
under intense, short ($T=1$\ldots20\,fs) 
pulses with a photon energy of
$E_\mathrm{ph}=12$\,keV focused to an area of
$A=(100\,\mathrm{nm})^2$.  The pulse  amplitude in time is given by 
\begin{equation}
  \label{eq:tdin}
  F(t)=\sqrt{I_\mathrm{peak}}\cos^2\left(c\frac{t}{T}\right)
  \quad\mbox{for } |t|\le (\pi T)/(2c)
\end{equation}
with $c=2\arccos(2^{-1/4})\approx1.144$ so that $T$ has the meaning  
of the full-width at half maximum pulse length of the intensity
$I(t)=F^2(t)$.  For a fixed number of $n=10^{12}$ photons in the pulse,
the peak intensity follows as  
\begin{equation}
  \label{eq:pkin}
  I_\mathrm{peak}=\frac{n\times E_\mathrm{ph}}{T\times A} 
  =\frac{1.92}{T/\mathrm{fs}}\:10^{22}\,\mathrm{W/cm}^2\,.
\end{equation}
Note that for different pulse lengths $T$, as discussed below,
the peak intensity $I_\mathrm{peak}$ changes. 
Our approach is based on a classical molecular dynamics
simulation similar to the one described in
\cite{saro02,misa+08}, with photo-ionization and Auger decay
treated by quantum-mechanical rates. 
The important effect of field-ionization is included by always
propagating the least bound electron of each atom/ion as a
classical particle in the field of all other ions and electrons.  
An ionization event is taken to have occurred, when this electron
leaves its mother ion beyond a threshold radius. We have found
the system dynamics to be insensitive to the particular choice
of threshold radius, so long as we stay within the natural
limits of half the nearest neighbor distance from above and a
minimum radius which avoids counting bound electrons with large
excursion radius as ionized, despite their immediate return to
the allowed region. Within this range we have settled on the
threshold radius value of 2.2 atomic units.  
Our approach has the advantage of incorporating the total field
of all charged particles, therefore including both
electron-impact ionization and ionization due to 
static fields \cite{gnsa+09}. 
We made use of an implementation
\cite{daka08} of the fast-multipole method \cite{grro87} to calculate
the Coulomb interaction of electrons and ions in order to bring
calculation times for the larger systems to a manageable level.

Paradigmatically, we first consider a Ne$_{1500}$ cluster under
a 10\,fs pulse. 
To clearly identify the effect of field ionization, we have
performed two calculations for each parameter set.
A full calculation including ionization from quasi-static
internal fields and a reduced calculation without this effect.  
The latter amounts in our approach to exclude the classical
propagation of a bound 2p electron with every atom/ion. 
At peak intensity, half-way through the pulse roughly each atom
will have been singly photo-ionized. 
However, with a radius of $R\approx20$\,{\AA} this cluster is not yet
large enough to capture any of the photo-electrons, cf.\
\fig{fig:elcapture}.  Without field-ionization this leads to a
homogeneous charging of the cluster, see \fig{fig:contours}c.  The
resulting electric field (\fig{fig:contours}a) exhibits a linear increase
inside the cluster and a $1/r$
dependence outside the cluster.  The maximum field at the surface is
about 1.5 atomic units which is sufficiently high for further
ionization and should not be neglected.

Field ionization generates many plasma electrons in
the cluster. More specifically,  at peak intensity there will already be an
average of two plasma electrons per cluster ion created in this
way.  Thereby, the cluster effectively becomes a charged conducting
sphere: The electrons neutralize the ionic charges at the center of
the cluster and the excess positive charge is localized on the surface
as can be seen in \fig{fig:contours}d in accordance with
phase~iii) described before.  As anticipated, this highly
efficient charge migration begins almost immediately with the
photo-induced charging of the cluster and takes place on a
sub-femtosecond time scale preceding the trapping of Auger electrons
as K-shell holes decay much more slowly.  The emergence of a neutral core
with a positively charged shell leads to a fundamentally different
electric field, as is shown in \fig{fig:contours}b: Although
the field at the surface is of similar strength as
in the homogeneously charged cluster, there is now a
nearly field-free region in the cluster center.  Within this region 
the ion motion is
therefore drastically suppressed at the expense of a violent explosion
of the more strongly charged outer shell.

The averaged radial trajectories of the surface ions starting at
$r\approx R$ and of the shell enclosing the inner half of the cluster
ions starting at $r\approx R/2^{1/3}$ (white lines in 
\fig{fig:contours}) indicate this twofold dynamics
of the cluster ions.  Clearly and in contrast to the reduced
calculation, the cluster core in the full calculation displays almost
no expansion throughout the pulse.

\begin{figure}[t]
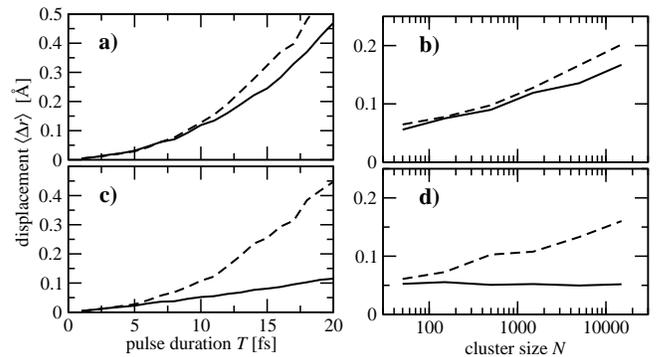

\includegraphics[scale=0.42]{Fig3a-displ-pulse.eps}
\includegraphics[scale=0.42]{Fig3b-displ-size.eps}
  \caption{Mean displacement of ions $\dr$ of neon clusters
  at peak intensity ($t=0$) for a full calculation (solid lines) and
  reduced calculations (dashed lines) neglecting field ionization. 
  Dependence on the pulse duration $T$ (left, for Ne$_{1500}$) and the
  cluster size $N$ (right, $T=10\,$fs), respectively. 
  The average $\dr$ is shown for  all cluster ions in the upper 
  panels and for only the inner half of the cluster ions in the lower 
  panels.}
\label{fig:displ}
\end{figure}%

The mean displacement 
\begin{equation}\label{DN}
\dr=\frac{1}{N}\sum_{i=1}^N
\left|\vec{r}_i(-\infty)-\vec{r}_i(0)\right|
\end{equation}
of the ions from their initial positions ($t\to-\infty$) provides a
systematic and more quantitative measure of the effect of charge
migration and its dependence on pulse duration $T$ and sample
size $N$.  It is
shown in \fig{fig:displ}a for various pulse lengths $T$ (but a fixed
photon number per pulse) in the full and the reduced calculation at
$t=0$, the time of peak intensity from which the largest contribution
to the diffraction pattern in an imaging experiment can be expected.
We find a strong amplification of ionic motion with increasing pulse
length.
Two factors contribute to this increase.  Firstly, a longer pulse obviously
means a longer propagation of an ion with its acquired momentum, and
secondly the charging of the cluster is higher for the longer pulses
due to the inherent time scale of about 2.5\,fs set by Auger decay.
Although the constant total photon number $n$ means that at peak
intensity each atom will have undergone roughly one photo-ionization
event for all pulse lengths, clusters under longer pulses will
additionally have seen Auger decays. Many of these Auger electrons
will escape the cluster, cf.\ \fig{fig:elcapture}, leading to a higher
charging.
The induced ion motion depends also on the sample size and shows a
strong increase of the mean displacement with increasing cluster size
(\fig{fig:displ}b).
Due to the additive
nature of the Coulomb force a larger cluster has  larger
electrostatic energy density than a smaller cluster of equal charge
density.  This leads to stronger ionic motion in larger clusters.

\begin{figure}[b]
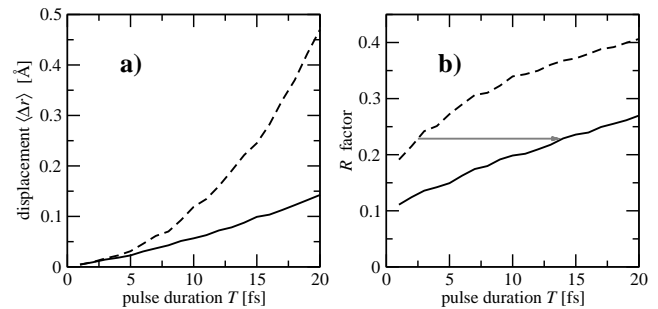

  \includegraphics[scale=0.4]{Fig4a-displ-tamper.eps}
  \includegraphics[scale=0.4]{Fig4b-factor-tamper.eps}
  \caption{Mean displacement (a) and $R$ factor (b) of
    ions of $\text{Ne}_{1500}$ cluster 
    embedded in He$_{15000}$ droplet (full line) and without
    helium droplet (dashed line) for varying pulse lengths.
    The gray arrow indicates a typical improvement in the damage
    tolerance, see text.} 
  \label{fig:tamper}%
\end{figure}%
As expected, the displacement is up to 20\% lower  (\fig{fig:displ}a) in the full
calculation, since the electrostatic energy of a charged conductor
(with the charge localized on the surface) is smaller than that of a
homogeneously charged sphere of equal charge.  
However, the displacement averaged over all ion positions does not
reflect the core-shell nature of the cluster ions with the almost
static core and the exploding shell in the case of effective field
ionization.  Therefore, \fig{fig:displ}c and \fig{fig:displ}d show the
displacement for the ionic core (the inner half of the ions) only,
revealing a dramatic effect: The mean displacement of the cluster core
atoms is reduced by up to 75\% for $\text{Ne}_{1500}$ for pulses of
20\,fs (\fig{fig:displ}c) and similarly 
for the larger cluster $\text{Ne}_{15000}$ at 10\,fs (\fig{fig:displ}d) 
length.
Yet, there seems to be a lower
limit for $\dr$ as is most obvious in \fig{fig:displ}d, but also
appears through the almost linear increase with pulse length $T$ in
\fig{fig:displ}c.  
This lower limit is due to the recoil acquired by the neon ions
during absorption of the 12\,keV photon, which is about
0.017\,\AA/fs. 

The different ion dynamics, split into an inert core and an exploding
shell, naturally leads to the idea of a sacrificial layer for imaging
experiments as proposed before \cite{halo+07} and further investigated in 
\cite{jufa08} for carbon-based samples.  Due to the different
material used the results are not directly comparable. However,
the actual dynamics of an exploding shell may be very different
in our case since field-ionized electrons are created much
faster than those due to electron-impact ionization from trapped
Auger-electrons. We consider here a 
Ne$_{1500}$ cluster embedded in a helium droplet of 15000 atoms.
During its buildup, the net
positive charge is efficiently and quickly transferred from the
cluster to the droplet.  The entire neon cluster now forms the core of
the composite system and remains mostly unscathed, while the helium
droplet, taking the role of the ionized shell, explodes as can be seen
in \fig{fig:tamper}a where $\dr$ is shown for the case with and
without the helium droplet.

In the context of coherent diffractive imaging it is important to ascertain 
that the strong reduction in $\dr$ translates into an equally improved quality
of the obtained diffraction pattern. We used our cluster
data to calculate the diffraction pattern corresponding to a spatial
resolution of 2\AA, where as in previous investigations \cite{halo+07} we
neglected contributions from the tamper and the plasma electrons. This
diffraction pattern, characterized by the intensity
$I_i^{\text{real}}$ registered at pixel $i$ 
of a detector having $k$ pixels, is then compared to an ideal diffraction
pattern $I_i^{\text{ideal}}$ without radiation damage.
We define
\begin{equation}
  \label{eq:rfac}
  R=\sum_{i=1}^k\left|\sqrt{I_i^{\text{real}}}
    -\sqrt{I_i^{\text{ideal}}}\right|\bigg/\sum_{j=1}^k
  \sqrt{I_j^{\text{ideal}}}\,,  
\end{equation}
which differs from a previous definition \cite{newo+00} in the normalization of
$\sqrt{I_i^{\text{real}}}$. 
The $R$ factor of \eq{eq:rfac} therefore measures discrepancies due to
electron loss (either homogeneous or inhomogeneous) and atomic motion.
Both influence crucially the image quality which can be obtained with
XFEL pulses. 
In comparison to the pure cluster we find a substantial
reduction of the $R$~factor for the helium embedded cluster
(\fig{fig:tamper}b).
Much longer pulses are tolerated by the embedded system until a
similar level of damage is reached as in the pure cluster. 
For example, an increase in pulse length by a factor of more
than five (from $T=2.5$\,fs to $T=14$\,fs, 
cf.\ gray arrow in \fig{fig:tamper}b)
becomes possible by embedding the cluster in helium.

We have presented detailed investigations for smaller clusters which
have revealed the importance of internal field ionization for
subsequent charge migration.  
For larger clusters ($R \gtrsim 60$\,{\AA}) trapping of 
photo-electrons becomes possible
and for clusters with radii in the hundreds of {\AA}ngstr{\"o}m
range, most photo-electrons will be trapped.  
Photo-electron trapping limits the 
average charge per atom in the cluster and the general rule of the
surface field-strength linearly increasing with cluster size is no
longer valid.  A simple estimate using the Bethe-rule for this
scenario nonetheless still predicts appreciable ionization through
quasi-static internal fields.  
For an $R=250$\,{\AA} cluster, for example, in
which 50\% of the atoms are singly photo-ionized, despite the
trapping of most photo-electrons, the surface-field will still
be sufficient for double ionization of a neutral atom. 
Hence, the field-ionization induced, ultrafast
charge migration can be expected to play an important role for
the success of single-shot coherent diffraction imaging
experiments for a wide array of the samples of interest.


\begin{thebibliography}{10}

\bibitem{fear+05}
J. Feld\-haus, J. Arthur, and J.~B. Hastings, J. Phys. B {\bf 38},  S\,799
  (2005).

\bibitem{gach07}
K.~J. Gaffney and H.~N. Chapman, Science {\bf 316},  1444  (2007).

\bibitem{newo+00}
R. Neutze, R. Wouts, D. van~der Spoel, E. Weckert, and J. Hajdu, Nature {\bf
  406},  752  (2000).

\bibitem{halo+05}
S.~P. Hau-Riege, R.~A. London, G. Huldt, and H.~N. Chap\-man, Phys. Rev. E {\bf
  71},  061919  (2005).

\bibitem{miho+01}
J. Miao, K.~O. Hodgson, and D. Sayre, Proc. Natl. Acad. Sci. U. S. A. {\bf 98},
   6641  (2001).

\bibitem{saro02}
U. Saalmann and J.~M. Rost, Phys. Rev. Lett. {\bf 89},  143401  (2002).

\bibitem{jufa+04}
Z. Jurek, G. Faigel, and M. Tegze, Eur. Phys. J. D {\bf 29},  217  (2004).

\bibitem{juos+04}
Z. Jurek, G. Oszlanyi, and G. Faigel, Europhys. Lett. {\bf 65},  491  (2004).

\bibitem{halo+04}
S.~P. Hau-Riege, R.~A. London, and A. Sz{\"o}ke, Phys. Rev. E {\bf 69},  051906
   (2004).

\bibitem{halo+07}
S.~P. Hau-Riege, R.~A. London, H.~N. Chapman, A. Sz{\"o}ke, and N.
  T{\^\i}mneanu, Phys. Rev. Lett. {\bf 98},  198302  (2007).

\bibitem{hach08}
S.~P. Hau-Riege and H.~N. Chapman, Phys. Rev. E {\bf 77},  041902  (2008).

\bibitem{krsm02}
V.~P. Krainov and M.~B. Smirnov, Phys. Rep. {\bf 370},  237  (2002).

\bibitem{sasi+06}
U. Saalmann, C. Siedschlag, and J.~M. Rost, J. Phys. B {\bf 39},  R\,39
  (2006).

\bibitem{siro04}
C. Siedschlag and J.~M. Rost, Phys. Rev. Lett. {\bf 93},  043402  (2004).

\bibitem{hobo+08}
M. Hoener, C. Bostedt, H. Thomas, L. Landt, E. Eremina, H. Wabnitz, T.
  Laarmann, R. Treusch, A.~R.~B. de~Castro, and T. M\"{o}ller, J. Phys. B {\bf
  41},  181001  (2008).

\bibitem{besa57}
H.~A. Bethe and E. Salpeter, {\em Quantum Mechanics of One- and Two-Electron
  Atoms} (Springer, Berlin, 1957).

\bibitem{misa+08}
A. Mikaberidze, U. Saalmann, and J.~M. Rost, Phys. Rev. A {\bf 77},  041201
  (2008).

\bibitem{gnsa+09}
C. Gnodtke, U. Saalmann, and J.~M. Rost, unpublished, 2009.

\bibitem{daka08}
H. Dachsel and I. Kabadshow, \\ http://www.fz-juelich.de/jsc/fmm, 2008.

\bibitem{grro87}
L. Greengard and V. Rokhlin, J. Comput. Phys. {\bf 73},  325  (1987).

\bibitem{jufa08}
Z. Jurek and G. Faigel, Eur. Phys. J. D {\bf 50}, 35 (2008).
\end{thebibliography}
\end{document}